\documentstyle[11pt,aaspp4,flushrt,psfig2]{article}

\slugcomment{Accepted for publication in ApJ}
\lefthead{M.~Kramer et al.}
\righthead{Profile instabilities of the millisecond pulsar PSR J1022+1001}

\begin{document}

\title{Profile instabilities of the millisecond pulsar PSR J1022+1001}

\author{Michael Kramer\altaffilmark{1,2}, 
Kiriaki M. Xilouris\altaffilmark{3},
Fernando Camilo \altaffilmark{4},
David J. Nice\altaffilmark{5},
Donald C. Backer\altaffilmark{1},
Christoph Lange\altaffilmark{2},
Duncan R. Lorimer\altaffilmark{2,3},
Oleg Doroshenko\altaffilmark{6},
Shauna Sallmen\altaffilmark{1}
}
\altaffiltext{1}{Astronomy Department, University of California, Berkeley,
CA 94720, USA}
\altaffiltext{2}{Max-Planck-Institut f\"ur Radioastronomie, Auf dem H\"ugel 69,
53121  Bonn, Germany} 
\altaffiltext{3}{National Astronomy and Ionosphere Center, Arecibo Observatory, P.O. Box 995, Arecibo, PR 00613, USA}
\altaffiltext{4}{Nuffield Radio Astronomy Laboratories,
Jodrell Bank, Macclesfield, Cheshire SK11 9DL, England; Marie Curie Fellow} 
\altaffiltext{5}{Joseph Henry Laboratories and Physics Department,
Princeton University, Princeton, NJ 08544, USA}
\altaffiltext{6}{Astro Space Center of P.N.~Lebedev Physical Institute,
Academy of Science, Leninski pr.~53, Moscow 117924, Russia }

\begin{abstract} 

We present evidence that the integrated profiles of some millisecond
pulsars exhibit severe changes that are inconsistent with the moding
phenomenon as known from slowly rotating pulsars. We study these profile
instabilities in particular for PSR J1022+1001 and show that they
occur smoothly, exhibiting longer time constants than those associated
with moding. In addition, the profile changes of
this pulsar seem to be associated with a relatively narrow-band
variation of the pulse shape. Only parts of the integrated profile
participate in this process which suggests that the origin of this
phenomenon is intrinsic to the pulsar magnetosphere and unrelated to
the interstellar medium. A polarization study rules out profile
changes due to geometrical effects produced by any sort of precession.
However, changes are observed in the circularly polarized radiation
component.  In total we identify four recycled pulsars which also
exhibit instabilities in the total power or polarization profiles due
to an unknown phenomenon (PSRs J1022+1001, J1730$-$2304, B1821$-$24,
J2145$-$0750).

The consequences for high precision pulsar timing are discussed in
view of the standard assumption that the integrated profiles of
millisecond pulsars are stable. As a result we present a new method to
determine pulse times-of-arrival that involves an adjustment of
relative component amplitudes of the template profile. Applying this
method to PSR J1022+1001, we obtain an improved timing solution with a
proper motion measurement of $-17\pm2$ mas/yr in ecliptic longitude.
Assuming a distance to the pulsar as inferred from the dispersion
measure this corresponds to an one-dimensional space velocity of 50 km
s$^{-1}$.

\end{abstract}
\keywords{Pulsars: millisecond pulsars -- normal pulsars -- profile stability --
precession -- mode changing -- timing precision -- PSRs J1022+1001,
1730$-$2304, B1821$-$24, J2145$-$0750 } 

\section{Introduction}

The discovery of millisecond pulsars (MSPs, \cite{bkh+82}), opened new
ways to study the emission mechanism of pulsars. Although pulse
periods and surface magnetic fields are several orders of magnitude
smaller than those of slowly rotating ('normal') pulsars, the emission
patterns show some remarkable similarities (\cite{kxl+98}, hereafter
KXL98; \cite{xkj+98}, hereafter XKJ98; \cite{jak+98}).  This suggests
that the same emission process might work in both types of objects
despite orders of magnitude difference in the size of their
magnetospheres. A systematic study of the emission properties of MSPs
and a comparison of the results with characteristics of normal pulsars
can thus lead to new important insight in the emission physics of
pulsars.  In this paper we investigate unexpected profile
instabilities seen for some MSPs and compare them to profile changes
known for normal pulsars.

The profile changes described here have consequences for high precision
timing of MSPs, in which integrated profiles are cross-correlated with a
standard template to measure pulse times of arrival.  This procedure
implicitly assumes that the shape of the integrated profile does not
vary.  This premise has never been tested thoroughly for MSPs.  The only
systematic search we are aware of was an analysis of the planet pulsar
PSR B1257+12 which was searched, unsuccessfully, for shape changes
(\cite{kw93}).

Recently, Backer \& Sallmen (1997) \nocite{bs97} noticed a significant
change in the profile of the isolated millisecond pulsar PSR B1821$-$24 in
about 25\% of all observations on time scales of a few hours and
possibly days.  Even earlier Camilo (1995) \nocite{cam95a} described
observations of PSR J1022+1001 which show profile changes on
apparently shorter time scales, i.e.~hours or less. These are the
first cases in which such instabilities of profiles averaged 
over many pulse periods have been reported.

The plan of the paper is as follows. After briefly reviewing what is known
about the profile stability of normal pulsars in the next section, we
present observations of PSR J1022+1001 and investigate a large data
set with respect to pulse shape changes in time, frequency and
polarization.  In Sect.~\ref{timing} we study the consequences for
high precision timing and present a method to compensate for the
profile changes when measuring pulse times-of-arrival. Possible
explanations for the observed profile changes are discussed in
Sect.~\ref{discus}, in view of additional sources showing a similar
phenomenon. A summary of this work is made in Sect.~\ref{summary}.

\section{Profile stability of normal pulsars}

\label{normal}

Since the discovery of pulsars it has been known that individual
pulses are highly variable in shape and intensity.  Nevertheless,
summing a sufficiently large number of pulses leads generally to a
very stable pulse profile.  Systematic studies of the stability of
integrated profiles were carried out by Helfand, Manchester \& Taylor
(1975)\nocite{hmt75} and recently by Rankin \& Rathnasree
(1995)\nocite{rr95}. These studies show that a number of a few
thousand pulses added together are very often enough to produce a
final stable waveform which does not differ from a high
signal-to-noise ratio (S/N) template by more than 0.1\% or even less.

Despite this stability of pulse profiles, a small sample of normal
pulsars shows distinct pulse shape changes on time scales of minutes.
This behaviour was first noticed by Backer (1970) \nocite{bac70a} and
is nowadays well known as {\em mode changing}.  In a mode change the
pulsar switches from one stable profile to another on a time scale of
less than a pulse period, remains in that mode for typically hundreds
of periods, before it returns back to the original pulse shape or
switches to another mode. This immediate switch from one mode to the
next is a common phenomenon and is often associated with a sudden
change in pulse intensity (\cite{ran86}). Interestingly, Suleymanova,
Izvekova \& Rankin (1996)\nocite{sir96} report that a mode switch in
PSR B0943+10 is preceded by a decline in intensity for one mode,
although again a so called ``burst'' and ``quiet'' mode can be
distinguished.

Rankin (1986)\nocite{ran86} noted that mode-changing is often observed
for such sources which exhibit rather complex profiles showing both
the so-called ``cone'' and ``core'' components (cf.~\cite{ran83};
\cite{lm88}).  The mode changing manifests itself as a reorganization
of core and cone emission and thus usually affects the {\em whole
profile} (often including polarization properties) rather than only
certain pulse longitudes. A rare counter-example might be PSR
J0538+2817 (\cite{acj+96}). 

A phenomenon related to mode changing might be {\em nulling}, i.e.~the
absence of any pulsed emission for a certain number of periods.  There
are no clear and unequivocal explanations for the origin of nulling
or mode changing which is normally interpreted as a re-arrangement in
the structure of the emitting region. Some studies report a possible
relationship between mode changing and change in emission height
(e.g.~\cite{bmsh82}). Other interpretations invoke a large variation
in the absorption properties of the magnetosphere above the polar cap
(e.g.~\cite{zqlh97}). In any case, mode changing will increase the
number of pulses that must be added before reaching a final
waveform. However, even for pulsars which show mode changes a maximum
number of $\sim10^4$ pulses is typically sufficient for a stable
average pulse shape to emerge from the process of adding seemingly
random pulses (\cite{hmt75}; \cite{rr95}). In contrast, the profile
changes of PSR J1022+1001 which we study in this paper, are on
much longer time scales, i.e.~hundreds of thousands of periods or more
as discussed below.

\section{The changing profile of PSR J1022+1001}

\label{1022}

\begin{figure}
\figurenum{1}
\centerline{\psfig{figure=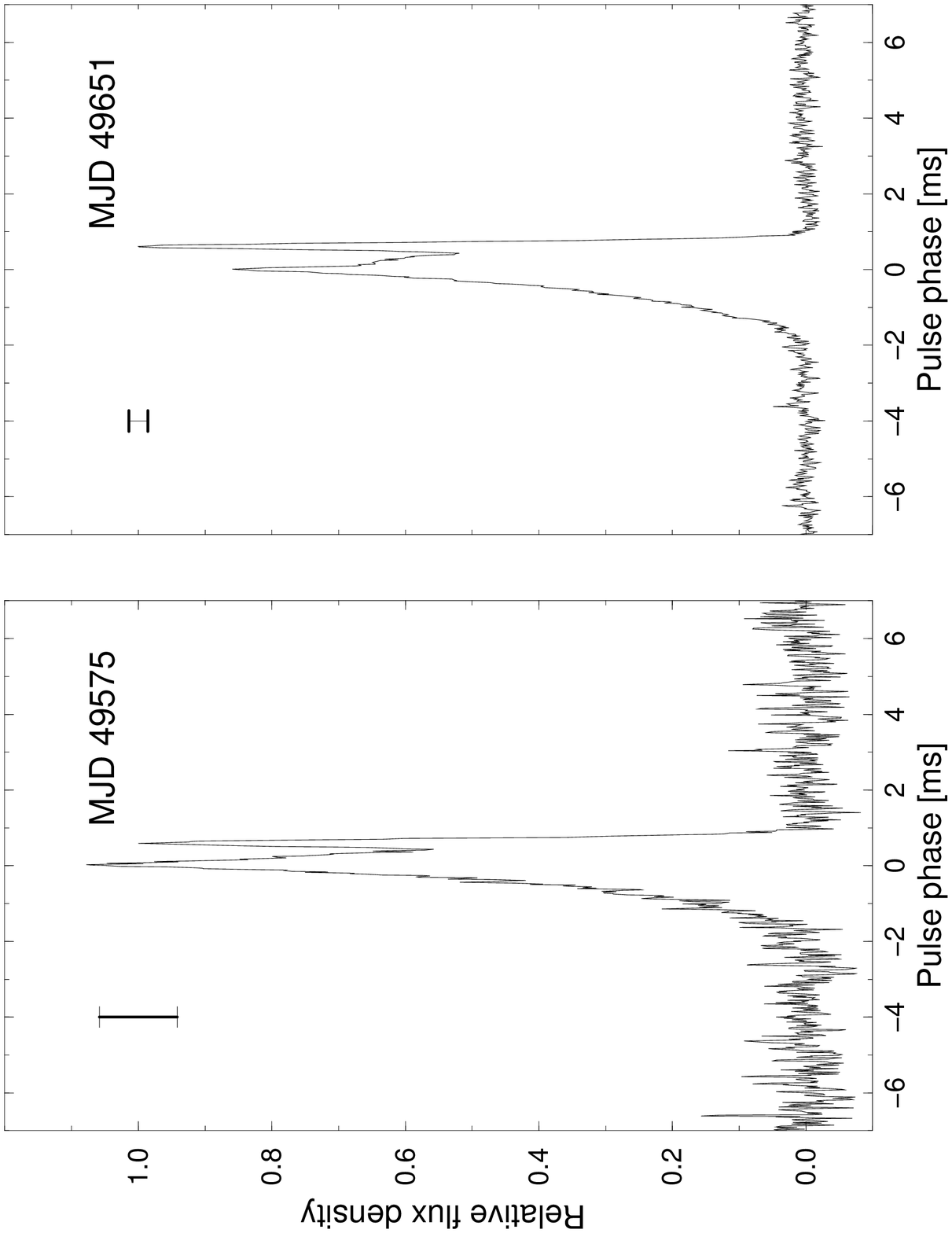,height=12cm,rotate=90}}

\bigskip

\figcaption{\label{first} Typical profile changes of PSR J1022+1001 as
seen during the first measurements obtained in Effelsberg at 1410
MHz. The error bars are conservative, being the result of $3\sigma$
values calculated from off-pulse data and a performed worst-case
analysis (see text for details). The profiles are significantly
different.}

\end{figure}

Soon after the discovery of PSR J1022+1001 (\cite{cnst96}), we
included it as part of our regular timing programme at Effelsberg.
The first high S/N profile was obtained in 1994 August
(Fig.~\ref{first}a). Comparison with another high S/N profile in 1994
October (Fig.~\ref{first}b) clearly demonstrated that the resolved
pulse peaks differ significantly in their relative amplitudes.
Observations at other telescopes confirmed this result which will be
discussed in detail in the following.

\subsection{Observations and data reduction}

The majority of the data presented in this paper were obtained at 1410
MHz with the Effelsberg 100-m radiotelescope of the
Max-Planck-Institut f\"ur Radioastronomie, Bonn, Germany. Besides the
Effelsberg Pulsar Observing System (EPOS) described by KXL98, we also
made measurements with the Effelsberg-Berkeley-Pulsar-Processor (EBPP)
--- a coherent de-disperser that has operating in parallel with EPOS
since 1996 October.

The EBPP provides 32 channels for each polarization with a total
bandwidth of up to 112 MHz depending on observing frequency, dispersion
measure and number of Stokes parameters recorded.  For PSR J1022+1001 a
bandwidth of 56 MHz can be obtained when recording only the two
orthogonal (left- and right hand) circularly polarized signals (LHC and
RHC). In polarization mode, i.e.~also recording the polarization
cross-products, a bandwidth of 28 MHz can be used. Each channel is
coherently de-dispersed on-line (assuming a dispersion measure of
DM$=10.25$ pc\,cm$^{-3}$) and folded with the topocentric pulse period.
Individual sub-integrations typically last for 2 min, before they are
transferred to disk.  A more detailed description of the EBPP can be
found in Backer et al.~(1997)\nocite{bdz+97} and Kramer et
al.~(1999)\nocite{kll+99}. 

\label{instrumental}

\begin{figure}
\figurenum{2}
\epsscale{0.75}
\plotone{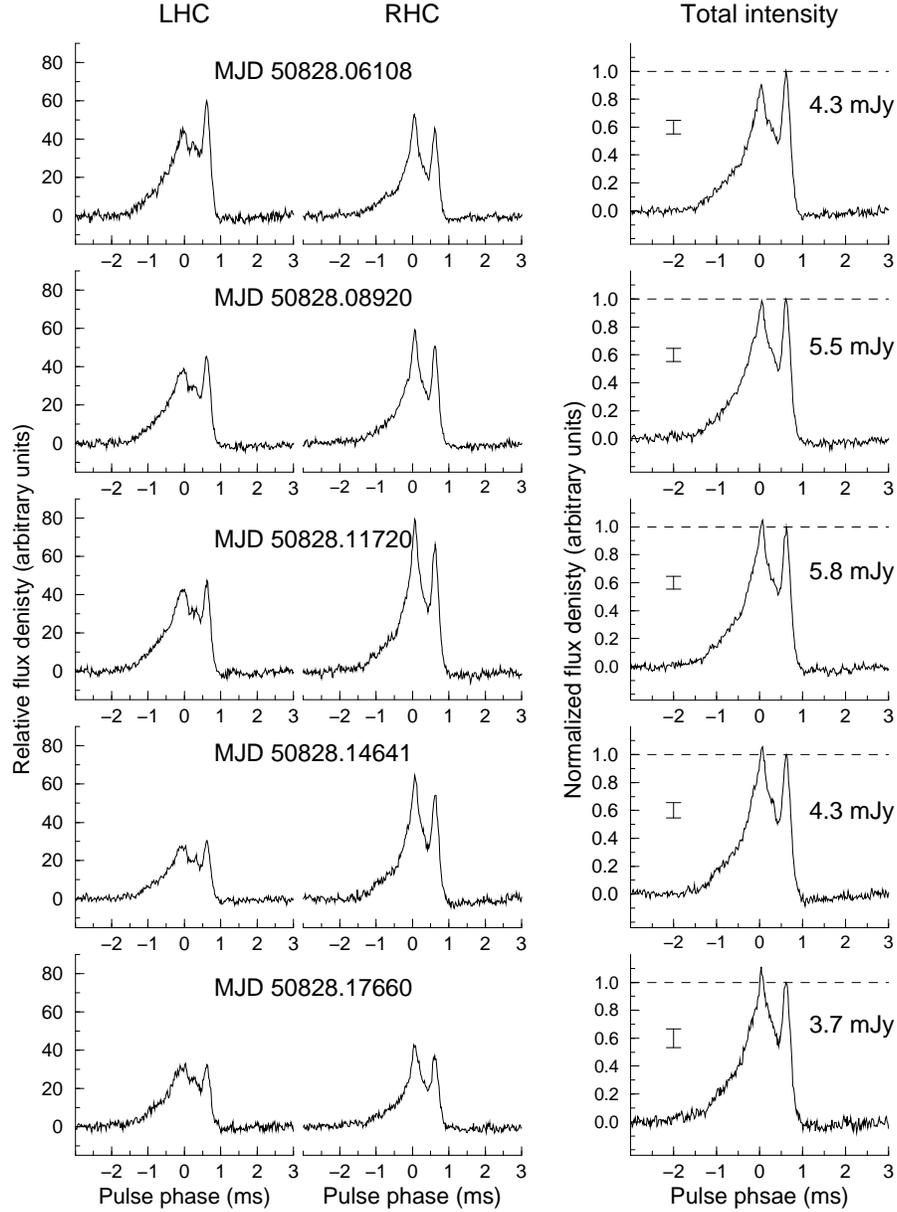}

\bigskip

\figcaption{\label{lhcrhc} 
Change of integrated profiles for PSR J1022+1001 
during a 3.3h observation showing
LHC (first column) and RHC (second column) signals and the total power
profile (third column) after correctly adding the two polarizations.
While LHC and RHC profiles are scaled to the same arbitrary flux units,
the total power profiles were scaled to unity amplitude of the
trailing pulse peak (dashed horizontal line). The error bars
are conservative, being the result of
a worst-case analysis (see text for details). Measured flux
densities noted for each corresponding profile demonstrate that the
profile changes are unrelated to pulse intensity.}

\end{figure}

In order to monitor the gain stability and polarization characteristics
of the observing system, we also performed regular calibration
measurements using a switchable noise diode. The signal from this noise
diode is injected into the waveguide following the antenna horn and was
itself compared to the flux density of known continuum calibrators
during regularly performed pointing observations.  Switching on the
noise diode regularly after observations of pulsars allowed monitoring of
gain differences in the LHC and RHC signal paths. Use of this procedure,
along with parallel observations by two independent data acquisition
systems, allows us to exclude an instrumental origin of the observed
profile changes. As a demonstration we show a three hour observation of
PSR J1022+1001 in Fig.~\ref{lhcrhc}, where each profile corresponds to
an integration time of about 40 min. The total power profile (right
column) was obtained after appropriately weighting and adding the LHC
and RHC profiles shown in the first two columns. In order to guide the
eye, we have drawn a dashed horizontal line at the amplitude of the
trailing pulse peak, which was normalized to unity. Error bars are based
on a {\em worst-case} analysis, combining $3\sigma$ values calculated
from off-pulse data with the (unlikely) assumption that the gain
difference has (still) an uncertainty of about 20\%. Inspecting the time
evolution of the shown profiles, we see that the RHC profile remains
unchanged during the whole measurement. At the same time the LHC profile
undergoes clear changes.  At the beginning of the observations, the
trailing pulse peak is the dominant feature in the LHC profile, it
then weakens gradually with time, until it becomes of equal amplitude to
the first pulse peak.  The resulting (total power) profiles reflect
exactly this trend. 

The measurement presented in Fig.~\ref{lhcrhc} clearly demonstrates that
the observed profile changes are not of instrumental origin, owing to the
lack of any instrumental effect which could explain the shown
evolution on the observed short time scales. Moreover, a correlation
between profile shape and source elevation or hour angle is not present.
We also searched for a possible relation between profile changes and
pulse intensity. In Fig.~\ref{lhcrhc} we thus indicate the flux density
measured for the corresponding profiles (with an estimated uncertainty of
less than 10\%).  Clearly, the profile changes are
uncorrelated with changes in intensity. Instead, the observed intensity
change is presumably caused by interstellar scintillation -- a common
phenomenon seen in low dispersion measure pulsars (\cite{ric70}).

\subsection{Profile changes with time}

\label{profilechanges}

\begin{figure}
\figurenum{3}
\plotone{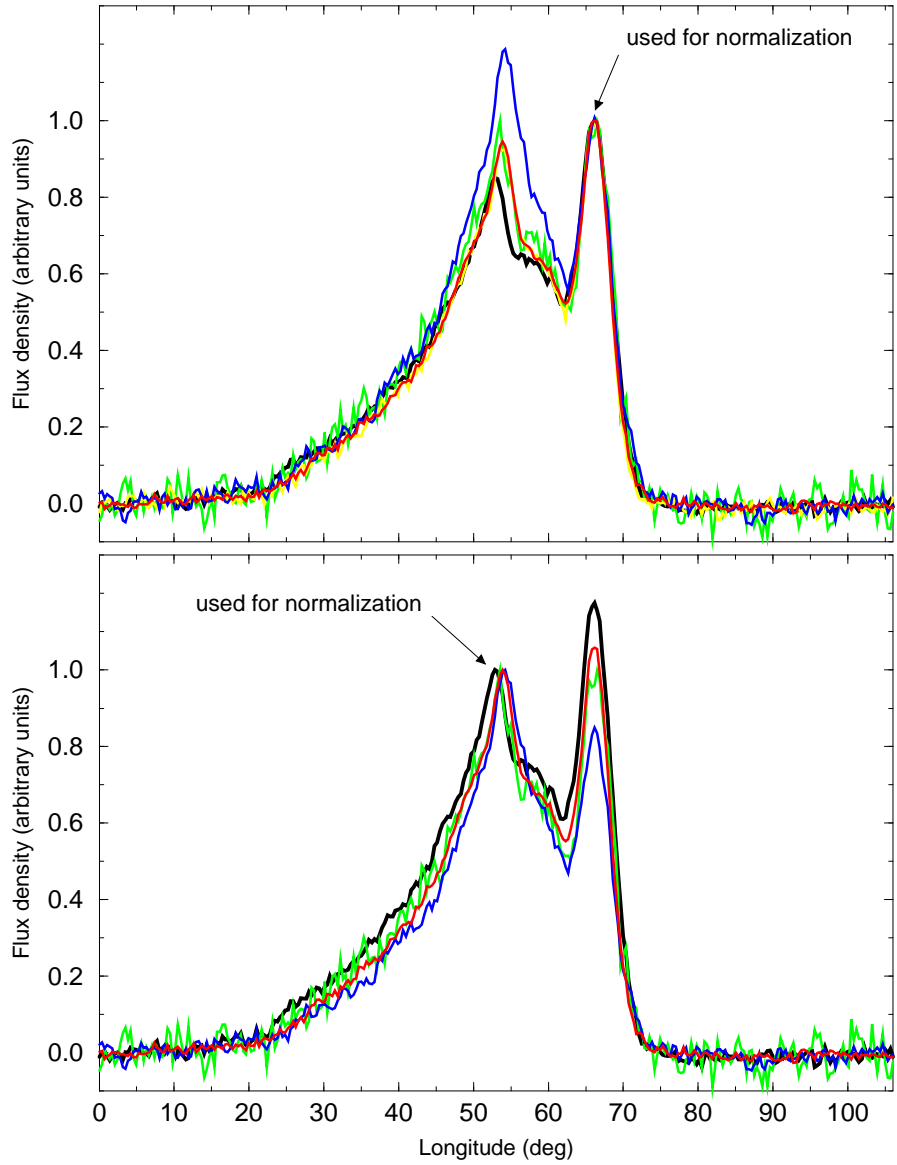} 

\bigskip

\figcaption{\label{diffnorm} 
Same set of pulse profiles of PSR J1022+1001 observed at 1410 MHz,
normalized to the leading (bottom panel) and trailing (top panel) 
pulse peak.}

\end{figure}

A simple comparison of measured pulse profiles normalized to each of the
two pulse peaks provides first clues as to whether the profile is
changing as a whole or stable parts are present.  In Fig.~\ref{diffnorm}
we present pulse profiles obtained at 1.410 MHz at different epochs.
Normalizing to the leading pulse peak, the profile apparently
changes over all pulse longitudes, i.e.~including the depth of the
saddle region and the width of the profile itself. Normalizing the same
profiles to the trailing pulse peak seems to cause mainly changes in the
first profile part while the trailing one remains stable.  The picture
seems also to apply to the 430-MHz data obtained by Camilo (1995) and
can be confirmed, as discussed later, by the timing behaviour of this
pulsar. 

\begin{figure}
\figurenum{4}
\epsscale{0.5}
\plotone{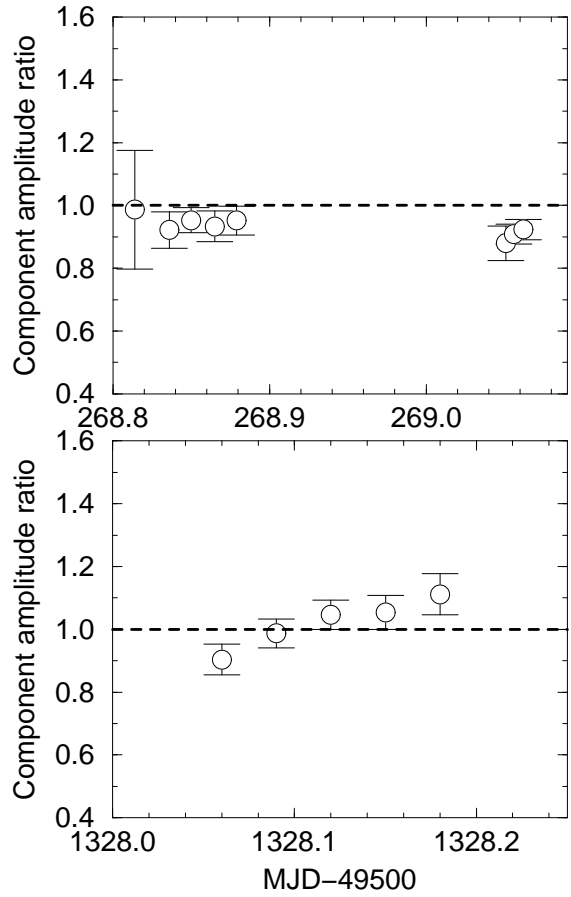}

\bigskip

\figcaption{\label{timescale} 
Ratio of pulse peak amplitudes (amplitude of leading peak divided by
that of trailing one) as a function of time for two different
observations of similar length. The difference in the time
scale of the profile changes is clearly visible.}

\end{figure}

Although our data sometimes suggest that variations on time scales of a
few minutes are present, we need higher S/N data to confirm this
impression. Instead, we reliably study here profile variations visible
on longer time scales by adding about $6\cdot10^4$ to $8\cdot10^4$
pulses each (i.e.~10 to 16 min). Although this corresponds to a much
larger number of pulses than needed to reach a stable profile for normal
pulsars even in the presence of moding (cf.~Sect.~\ref{normal}), we
still observe a {\em smoothly varying} set of pulse shapes. In order to
demonstrate that the involved time scales are highly variable, we
calculated the amplitude ratio of the leading and trailing pulse peaks
at 1410 MHz as the most easily accessible parameter to describe the
profile changes. In order to use a large homogeneous data set, we
analyzed EPOS data, which were obtained with a bandwidth of 40 MHz, a
time resolution of $25.8\mu$s (cf.~KXL98) and an integration time as
quoted above. We estimated uncertainties in the amplitude ratio using
the same worst-case analysis as described before. The mean value of the
component ratio (amplitude of the leading pulse peak divided by that of
the second one) for the whole data set covering about four years of
observations is $0.975\pm0.009$.  Two examples of observations of
comparable duration are shown in Fig.~\ref{timescale} where we plot the
amplitude ratio as a function of time. During the first measurement the
profile appears to be stable. In the second observation profile changes
are evident. This is consistent with the results of an unsuccessful
search for periodicities or typical time scales in the amplitude ratio
data by computing Lomb periodograms of the unequally sampled data set.
Using a method described by Press et al.~(1992)\nocite{ptvf92} we
investigated time scales ranging from several hours, over days to months
without obtaining significant results.

\begin{figure}
\figurenum{5}
\centerline{\psfig{figure=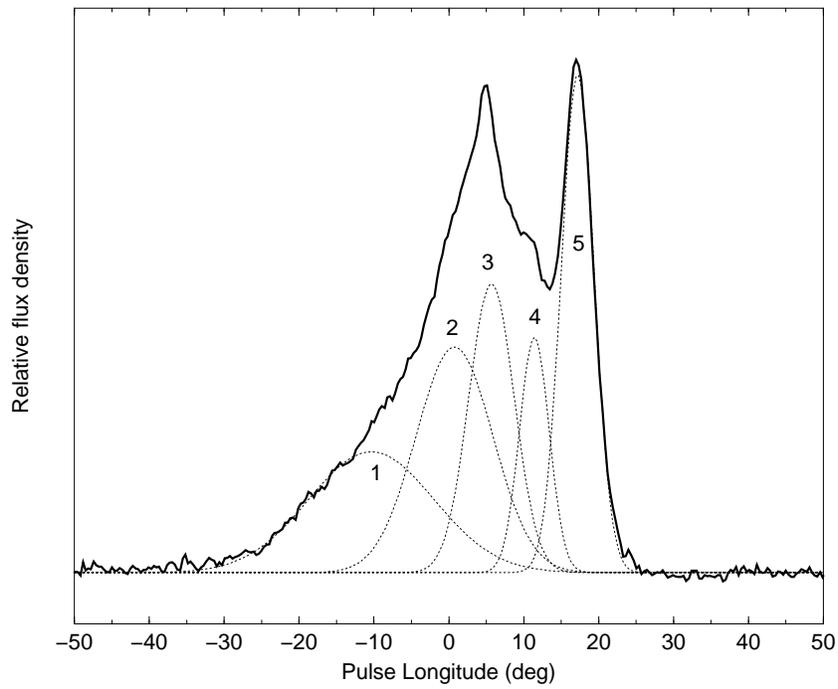,height=11cm,rotate=90}}

\bigskip

\figcaption{\label{compfit} 
Example of a profile observed for PSR 1022+1001 at 1410 MHz separated into
five Gaussian components. Adjusting only their relative
amplitudes describes all observed different pulse shapes with surprising
accuracy.}

\end{figure}

In order to model the profile changes in detail, we fit the integrated
profiles to a sum of Gaussian components, defined as
\begin{equation}
\label{gauss}
I(\phi) = \sum_{i=1}^{n} a_{3i-2} \;
          \exp \left\{ -\left( \frac{ \phi - a_{3i-1} - \phi_0 }{ a_{3i}}
          \right)^2 \right\},
\end{equation} 
where $\phi_0$ is a fiducial point. As shown by KXL98 PSR J1022+1001
is well described by a sum of $n=5$ components
(cf.~Fig.~\ref{compfit}).  We applied this method to the whole data
set, varying amplitude, positions, and widths of the components.  We
then developed a model using the median values of component position
and width, and found that, surprisingly, this model fits {\it all}
observed profiles well with only adjustments to the relative
amplitudes of the components.  Of the profiles studied, only 5\% of
the fits would have been rejected by the criteria of Kramer et
al. (1994)\nocite{kwj+94}. According to these, the significance level
of the null-hypothesis that the post-fit residuals in the on-pulse
region and the data in an off-pulse region of similar size are drawn from
the same parent-distribution, must not be less than 95\%. Most of the rare
cases, where these criteria were not fullfilled, were profiles with
very high S/N, indicating a refined model would be needed to perfectly
describe the best data.
 
\begin{figure}
\figurenum{6}
\centerline{\psfig{figure=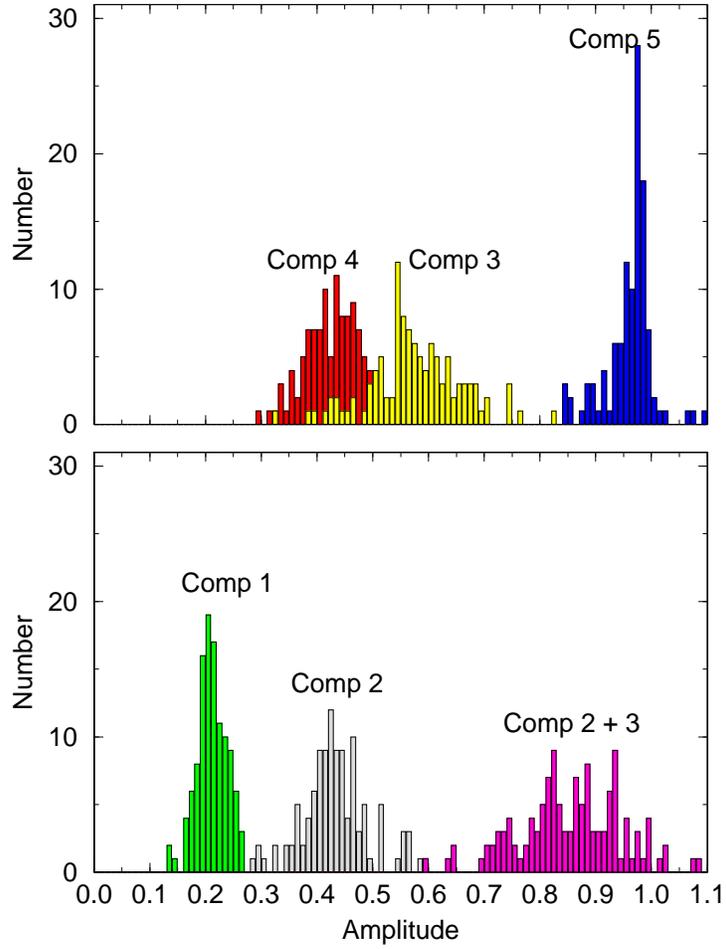,height=13cm}}

\bigskip

\figcaption{\label{amplhist} 
Statistical results of fitting a model of
five Gaussian components to the profiles observed at 1410 MHz (see
text). During these fits only the amplitudes of the Gaussians were
adjusted. Obtained amplitude distributions are shown (see Fig.~5 for
component numbering). The right distribution in the bottom panel
reflects an appropriate linear combination of the second and third
component, corresponding to the first pulse peak (see text).}

\end{figure}

\begin{figure}
\figurenum{7}
\centerline{\psfig{figure=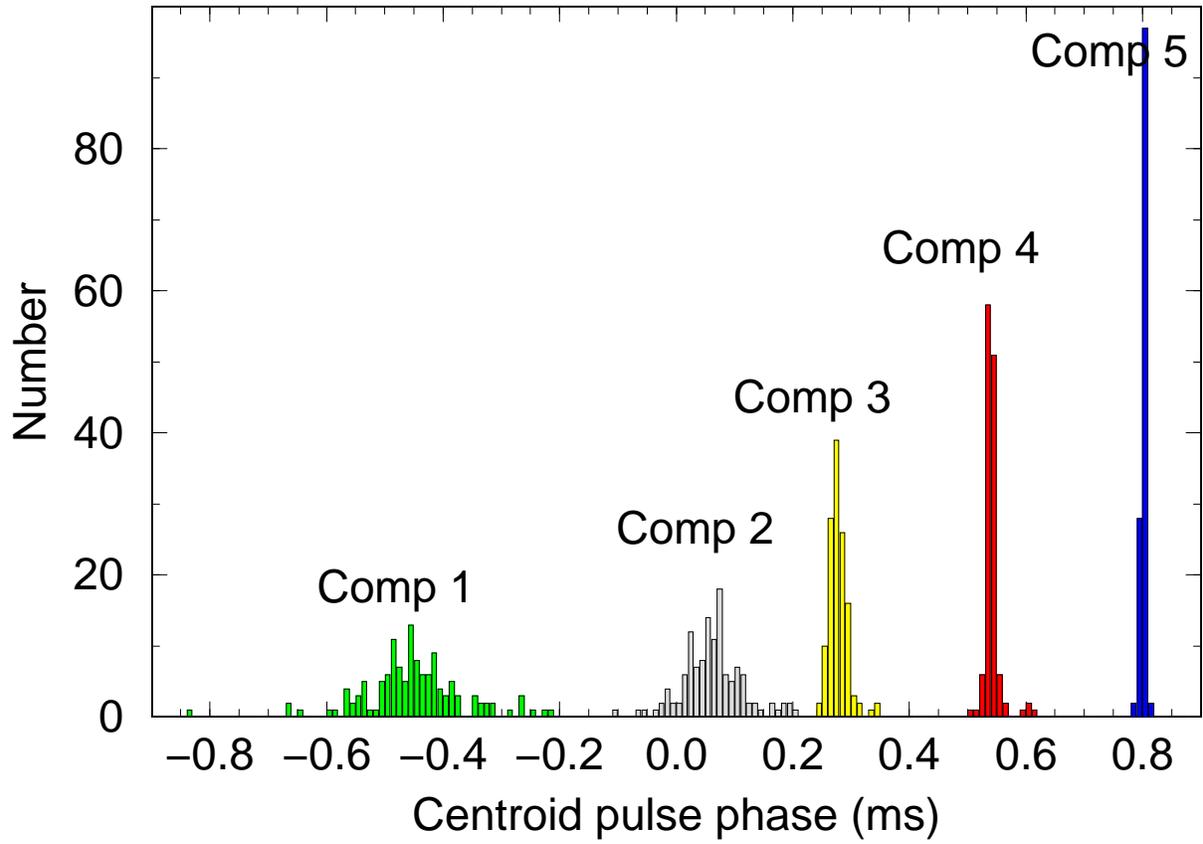,height=12cm}}

\bigskip

\figcaption{\label{centrhist} 
Statistical results of fitting a model of
five Gaussian components to the profiles observed at 1410 MHz (see
text). During these fits the amplitudes and centroids of the Gaussians were
adjusted. Obtained centroid distributions are shown (see Fig.~5 for
component numbering).}

\end{figure}

The summarized results of our Gaussian fitting procedure are presented
as a set of histograms in Figures~\ref{amplhist} and \ref{centrhist}
and Table~\ref{stattab}. Figure~\ref{amplhist} shows the occurrence of
amplitudes for each of the five components (for a numbering see
Fig.~\ref{compfit}). All profiles were normalized to the trailing peak
of the profile, so that the amplitude of the fifth component is always
close to unity. Whilst the range of amplitudes is well confined for
the first component, the amplitudes for the second and fourth
component show a broad distribution. In particular the amplitudes of
the third component exhibit a large scatter which is also demonstrated
by the summary of the results in Table~\ref{stattab}. Inspecting
Fig.~\ref{compfit}, it is clear that the amplitude, $p$, of the first
pulse peak is made up by a combination of intensities from component 2
and 3, scaling as $p=0.75a_4+0.95a_7$. The quantity $p$ is is also
displayed in Fig.~\ref{amplhist}, showing a very broad distribution
reflecting the observed changes in amplitude ratio.

We can now test numerically as to whether only parts of the leading
profile are changing by repeating the above analysis, but this time
allowing additionally for a fit in the relative spacing of the
components. The intriguing result is presented in Fig.~\ref{centrhist},
which shows the distributions of the resulting centroids (relative to the
fiducial point) for each component, and Table~\ref{stattab}.
Interestingly, the scatter in the central position gradually decreases
from the leading to the trailing part. At the same time, the
corresponding amplitude histogram is almost identical to
Fig.~\ref{amplhist} (not shown). This result indeed suggests that the
trailing part of the profile is more stable than the leading one, which
undergoes significant profile changes.

\subsection{Profile changes with frequency }
\label{freqdep}

Although profiles of normal pulsars are well known to change
significantly with observing frequency, MSPs show often a much smaller
profile development (XKJ98). In contrast, the profiles of PSR J1022+1001
show strong changes with frequency, which are inconsistent with the
canonical behaviour of normal pulsars (cf.~\cite{ran83},
\cite{lm88}). 

\subsubsection{Large frequency scale}

Comparing the {\em average} pulse profiles of PSR J1022+1001 over a
wide range of frequencies (cf.~\cite{snt97}, \cite{cnst96},
\cite{kkwj97}, \cite{sal98}, KXL98, Kramer et al.~1999) it becomes
clear that profile changes at frequencies other than 400 or 1400 MHz
are more difficult to recognize (but nevertheless possible).  Only
around these two frequencies both prominent pulse peaks are of
comparable (although nevertheless changing) amplitude. It has to be
addressed by simultaneous multi-frequency observations as to whether
the profile changes at different frequencies occur
simultaneously. This might, however, be a difficult task given the
discovered phenomenon discussed below.

\subsubsection{Small frequency scale }

\begin{figure}
\figurenum{8}
\centerline{\psfig{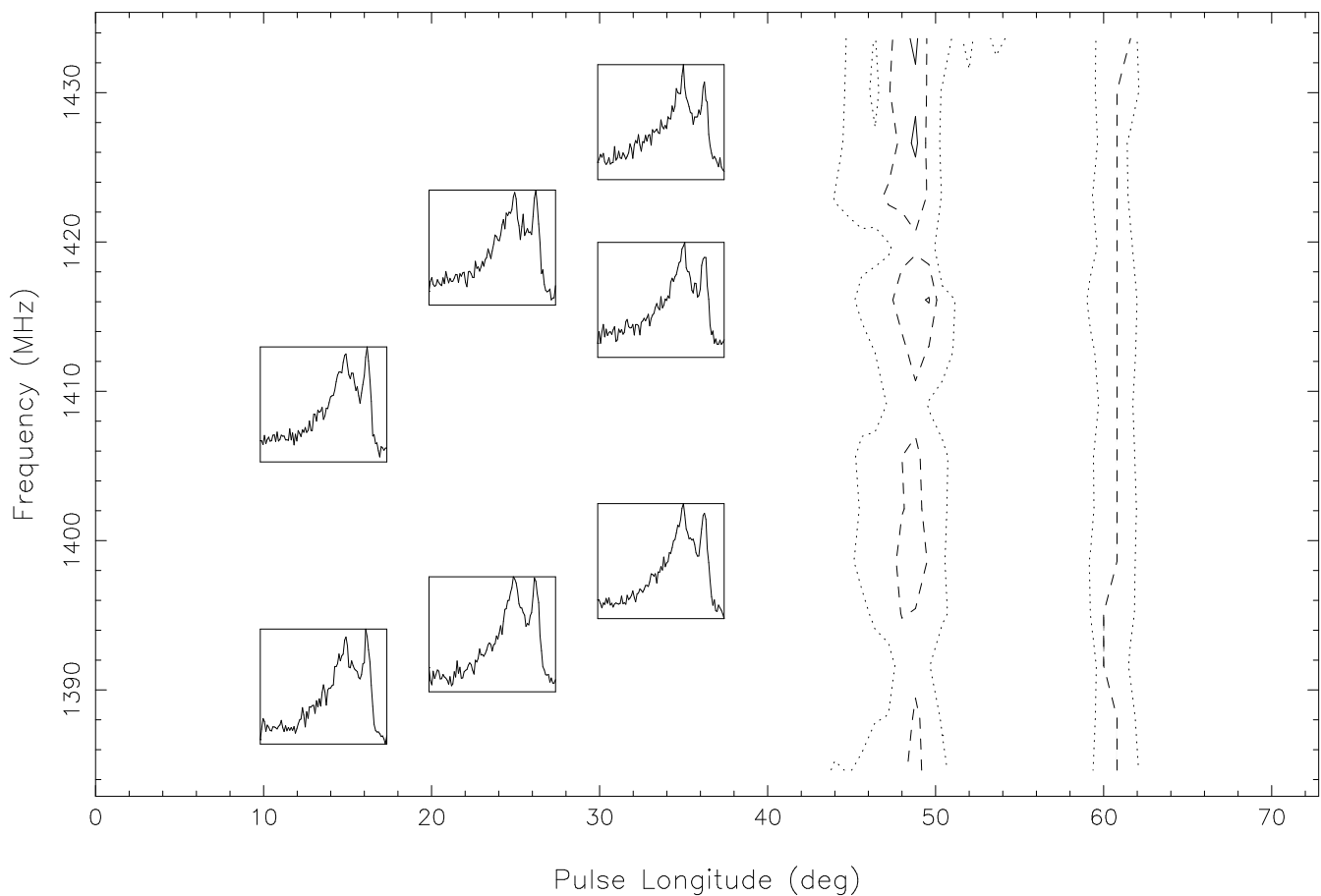}}

\bigskip

\figcaption{\label{bandpass} 
Contour plot of pulse intensity as a function of pulse longitude and
observing frequency covering a bandwidth of 52.5 MHz centered at
1410 MHz. Solid (dotted)
contour lines reflect an increase (decrease) of $3\sigma$ from the unit
amplitude of the trailing pulse peak (dashed contours).  A sample of
measured profiles is shown as insets whose vertical position corresponds
to their actual observing frequency. The longitude ranges of the contour
plot and shown pulse profiles are identical.  } 

\end{figure}

For almost all cases, EPOS and EBPP, both operating in parallel,
yielded identical pulse profiles. However, at a few occasions the EBPP
profiles differed slightly from those obtained with EPOS. The causes
are discovered profile variations across the observing bandpass: while
EPOS always uses a fixed bandwidth of 40 MHz, the bandwidth of the
EBPP for PSR J1022+1001 at 1410 MHz is 56 MHz in total power mode and
28 MHz in polarization mode. When profile changes happen on frequency
intervals smaller than this, the obtained profile depends on the exact
location and size of the bandwidth used. This is what we observe as
demonstrated by a contour plot (Fig.~\ref{bandpass}), where we show
the intensity as a function of pulse longitude and observing
frequency. In order to produce this plot we have added 12 min of EBPP
total power data folding with the topocentric pulse period. From 30
frequency channels (or 52.5 MHz, i.e. two channels were excised due to
technical reasons) each two adjacent ones were collapsed to produce a
reliable S/N ratio.  All resulting 15 profiles were normalized to the
second pulse peak, indicated by the dashed vertical line at 60$^\circ$
longitude. Contour levels were chosen such that solid lines reflect an
increase of $3\sigma$ (computed from off-pulse data) from the unit
amplitude of the trailing pulse peak.  Conversely, the dotted lines
denote $3\sigma$ decreases with respect to the trailing pulse
peak. Additionally, we overlay a sample of corresponding profiles as
insets whose vertical position reflects their actual observing
frequency. Their horizontal position is arbitrarily chosen for reasons
of clarity.  The longitude ranges covered in the contour plot and the
pulse profiles are identical.  Evidently, a significant profile change
is occurring on a small frequency scale of the order of 8 MHz, which
however also varies for different observations. 

Obviously, the profile observed over a large bandwidth is an average
of the individual profiles within the band. Depending on the relative
occurrence and strength of the various pulse shapes, which is
additionally modulated by interstellar scintillation, a whole variety
of pulse shapes and time scales can be created.

\subsection{Polarization structure}

\label{polsec}

\begin{figure}
\figurenum{9}
\centerline{\psfig{figure=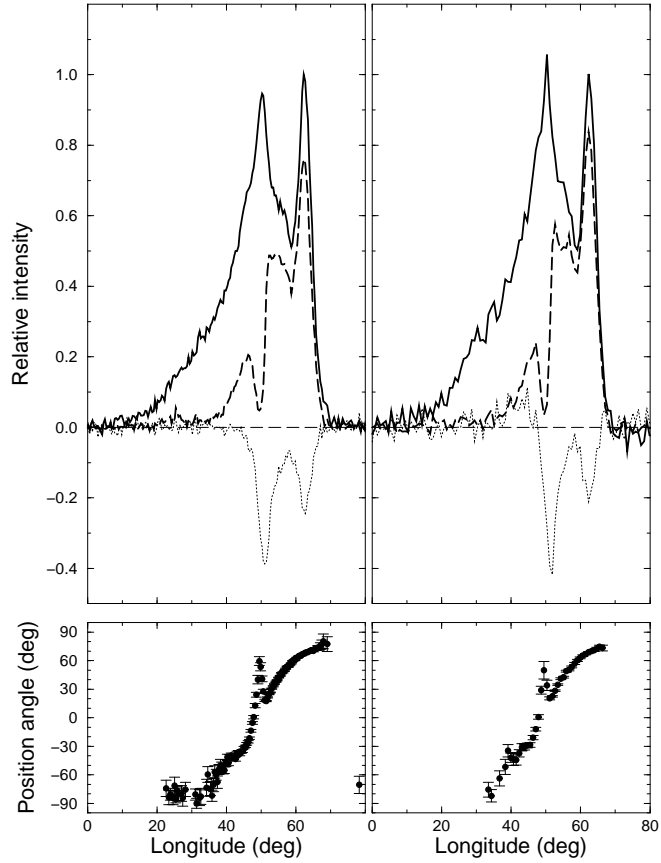,height=13cm}}

\bigskip

\figcaption{\label{poldata} 
Polarization characteristics observed for different
pulse shapes at 1410 MHz. While the linear polarization (dashed line)
is essentially unchanged, significant differences in circular
polarization (dotted line) like a sense reversal at position of 
the first pulse peak are visible in the right plot while missing 
on the left.}

\end{figure}

The polarization of PSR J1022+1001 has been already discussed by XKJ98
(see also \cite{sal98} and \cite{sta98}). Here we concentrate on the
impact of the profile changes on the polarization characteristics,
since it is already clear from Fig.~\ref{lhcrhc} that some changes are
to be expected.  In Fig.~\ref{poldata} we present polarization data
obtained with the EBPP at 1410 MHz for two typical pulse shapes.  In
the left panel the leading pulse peak is weaker, whereas in the right
panel the amplitude ratio is reversed.  The linearly polarized
intensity (and thus its position angle) is very similar in both
measurements, but the circular polarization shows distinct
differences. In the right profile we observe significant circular
power with positive sense, coinciding with the leading resolved peak
of the pulse profile. This feature of circular polarization is not
present in the left profile.  Similarly, the saddle region of the
right profile shows a dip in circular power, while at the same
longitude the left profile shows significant circular power with
negative sense.  Since the position angle swing appears to be
identical in both measurements, it rules out some obvious effects due to
changes in the viewing geometry. In fact, the strange notch appearing
at the maximum of circular power is prominent in both profiles and
seems to describe a {\em resolvable} jump by about $\sim70$ deg above
the otherwise fairly regular {\sf S}-like swing.  We stress, that the
two profiles shown in Fig.~\ref{poldata} represent only two typical
pulse profiles. Various states between two extremes can be observed.

The Gaussian components used to model the profile show a distinct
correspondence to the polarization structure: the first component
coincides with the unpolarized leading part of the profile. The second
component corresponds to the first linearly polarized feature, whilst
the third component resembles the first large peak in circular
polarization. The fourth component coincides with the second peak in
linearly polarized intensity, and the fifth Gaussian clearly agrees
with the trailing prominent pulse peak.  This correspondence between
Gaussian components and polarization features, along with the success
of the Gaussian model for the various profiles of this and other pulsars
(e.g.~\cite{kwj+94}, KXL98), strongly suggests that the Gaussian
components have some physical meaning, and are not just a mathematical
convenience used to describe profiles.

\subsection{Timing solution}

\label{timing}

We have undertaken timing observations of PSR~J1022+1001 at several
observatories and several observing frequencies over a span of four
years.  In many cases the timing measurements were derived from the same
data as used in the profile shape study described above.  Data were
collected at the 300\,m telescope at Arecibo\footnote{The Arecibo
Observatory, a facility of the National Astronomy and Ionosphere Center,
is operated by Cornell University under a cooperative agreement with the
National Science Foundation.} (May to November 1994; 430 MHz), the
100\,m telescope at Effelsberg (December 1994 to July 1998; 1400 MHz);
the 76\,m Lovell telescope at Jodrell Bank (April 1995 to July 1997; 600
and 1400 MHz); and the 42\,m telescope at Green Bank (July 1994 to May
1998; 370, 600, and 800 MHz).  At each observatory, data were folded
with the topocentric pulse period, de-dispersed (on- or off-line), and
recorded, along with the observation start time. 

\begin{figure}
\figurenum{10}
\centerline{\epsscale{0.6}\plotone{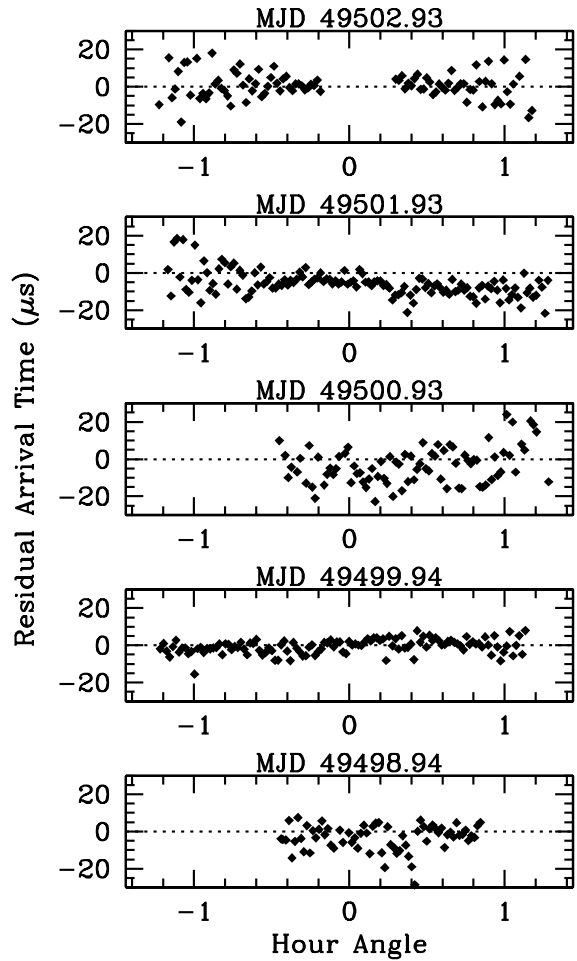}}

\bigskip

\figcaption{\label{resarecibo} Residual arrival times of the first five epochs
of Arecibo data.  Trends evident in the data include drifts in arrival 
times as well as overall offsets from zero.  As discussed in the text,
these artifacts are likely instrumental in nature.}
\end{figure}

Times of arrival were calculated by cross-correlating the data profiles
with a standard template.  For the Green Bank and Jodrell Bank data, a
template with fixed shape was used.  For the Arecibo and Effelsberg
data, the model of five Gaussian components with fixed width and
separation but freely varying amplitudes was used.  The Arecibo data
were not calibrated (left- and right-hand circular polarizations were
summed with arbitrary weights), and systematic trends were evident in
the residual arrival times, even after allowing Gaussian component
amplitudes to vary.  The trends were reduced somewhat by fitting the
residuals to a linear function of the amplitudes of the five Gaussian
components and removing the resulting function from the data.  These
procedures had the net effect of reducing the rms residual arrival times
from $25\,\mu$s to $17\,\mu$s for two minute integrations.  Still, some
systematics remained, typically drifts of order $20\,\mu$s over time
spans of 2 hours (Figure~\ref{resarecibo}). 

An alternative scheme for timing the Arecibo data, in which a
conventional fixed-template scheme was used, but only that part of the
profile from the central saddle point through the trailing edge were
given weight in the fit, gave results very similar to those of the
five-Gaussian fit.  We view this as further evidence that the trailing
edge of the profile is relatively stable, while the leading profile is
variable. 

\begin{figure}
\figurenum{11}
\centerline{ \plotone{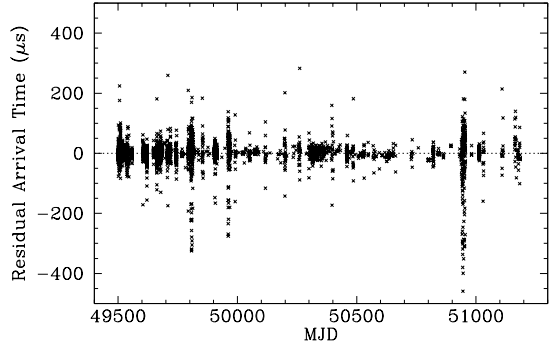}}

\bigskip

\figcaption{\label{resall} Residual arrival times of all data points.}
\end{figure}

A total of 4277 times of arrival (TOA) were measured.  These were fit to
a model of pulsar spin-down, astrometry, and orbital elements using the
{\sc tempo} program.  Root-mean-square (RMS) residual arrival times
after the fit were of order 15-20\,$\mu$s for the Arecibo, Jodrell Bank,
and Effelsberg data sets, and 40-100\,$\mu$s for the Green Bank data
(Figure~\ref{resall}).  To partially compensate for systematic
uncertainties, the Arecibo TOAs were given uniform weights in the fit
(equivalent to a timing uncertainty of 17\,$\mu$s), and systematic terms
(of order 10\,$\mu$s) were added in quadrature to the uncertainties of
TOAs from other observatories.  The resulting fits had reduced $\chi^2$
values close to 1 for each observatory, and the overall fit had a
reduced $\chi^2$ of 1.09 for the full data set.  Our best estimates of
timing parameters are listed in Table~\ref{timpar}.  To guard against
remaining systematic errors, we separately analyzed several subsets of
the data and incorporated the spread in parameters thus derived into the
uncertainties in Table~\ref{timpar}.  Particular data sets considered
included the individual sets from Green Bank, Jodrell Bank, and
Effelsberg; a smoothed data set (in which all TOAs from a given day were
averaged); and a data set which excluded all earth-pulsar lines-of-sight
which passed within $30^\circ$ of the Sun.  We recommend that the
uncertainties thus derived be treated as $1\sigma$ values. 

Because this pulsar is close to the ecliptic, the uncertainty in
ecliptic latitude, as determined by timing, is much greater than the
uncertainty in ecliptic longitude.  To minimize covariance between fit
parameters, the pulsar's position and proper motion are thus best presented in
ecliptic coordinates.  The ecliptic coordinates given in Table
\ref{timpar} are based on the reference frame of the {\sc DE 200}
ephemeris of the Jet Propulsion Laboratory, rotated by $23^\circ
26'21.4119''$ about the direction of the equinox. 

The proper motion of this pulsar has not been previously reported.  The
measured proper motion in ecliptic longitude, $\mu_\lambda$, translates to a
one-dimensional space motion of 50\,km\,s$^{-1}$, assuming a distance of
0.6\,kpc, as inferred from the dispersion measure.  This is typical of the
velocities of millisecond pulsars (e.g.~\cite{lor95}, \cite{cc98}).

\section{Discussion}

\label{discus}

\begin{figure}
\figurenum{12}
\centerline{\psfig{figure=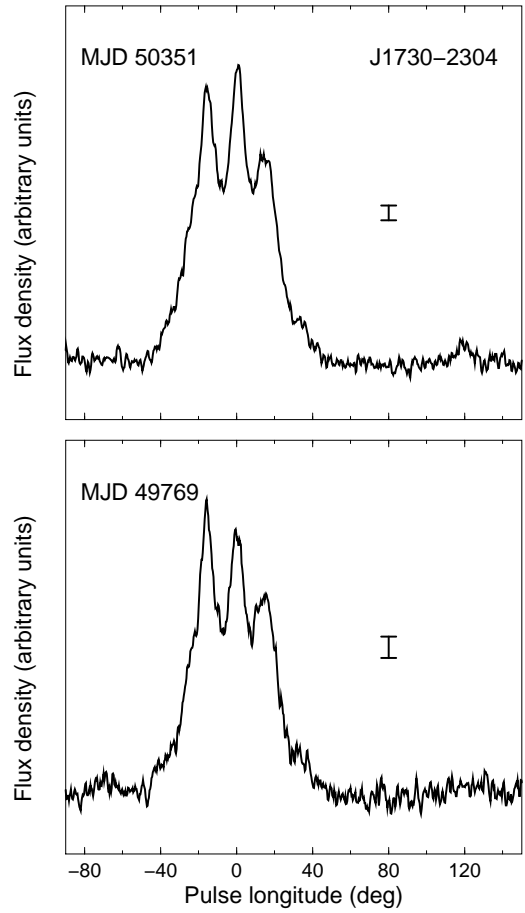,height=13cm}}

\bigskip

\figcaption{\label{psr1730} Two profiles of PSR J1730$-$2304 
measured in Effelsberg at 1410 MHz at different epochs. Error
bars reflect $3\sigma$ values calculated from off-pulse data.
A clear change in the profile is visible.}

\end{figure}

For PSR J1022+1001 we clearly demonstrated the existence of highly
unusual changes in pulse shape and polarization which cannot be
explained by instrumental effects.  Studies of other MSPs reveal that
PSR J1022+1001 is not the only source for which such behaviour can be
observed.  Backer \& Sallmen (1997) have already discussed a similar
phenomenon for PSR B1821$-$24. Another MSP where we find profile
changes is PSR J1730$-$2304 (see Fig.~\ref{psr1730}). Its usual
weakness at 1410 MHz (cf.~KXL98) prevents a data analysis as possible
for PSR J1022+1001, but similar profile changes have also been observed at
the Parkes telescope (Camilo et al.~in prep.).  Very recently,
Vivekanand, Ables \& McConnell (1998)\nocite{vam98} also described
small profile changes of PSR J0437$-$4715 at 327 MHz. Although they
observed this highly polarized pulsar only with a single polarization,
and although Sandhu et al.~(1997)\nocite{sbm+97} demonstrate that
measurements of this pulsar are difficult to calibrate, Vivekanand et
al.~argue that these pulse variations are real. In any case, the low
time resolution of their observed profiles prevents a detailed
analysis.

It was already noted by XKJ98 that for some MPSs profile changes can
be prominent in the polarization characteristics whereas the total
intensity remains mostly unchanged. As an intriguing example we refer
to PSR J2145$-$0750, for which XKJ98 measured at 1410 MHz a high
degree of polarization and a well defined, flat position angle (see
their Fig.~1). Recent results indicate that for most of the time, the
profile seems in fact to be weakly polarized with a highly disturbed
position angle swing (\cite{sal98}, \cite{sta98}).  However, a profile
very similar to XKJ98's 1410 MHz observation has been observed by Sallmen
(1998) also at 800 MHz.

As it is apparently the case for PSRs J1022+1001 and B1821$-$24, only
certain parts of the profile seem to actually change. Thus, we can
exclude any propagation effect due to the interstellar or
interplanetary medium since it should affect all parts of the profile
simultaneously.  When we compare the properties of this 'strange'
sample of MSPs, we notice that PSRs J0437$-$4715, J1022+1001 and
J2145$-$0750 have an orbiting companion while both PSRs J1730$-$2304
and B1821$-$24 are isolated pulsars. The existence of a binary
companion is therefore certainly unrelated to the observed phenomenon.
The pulse periods of the pulsars range from 3.05 ms (PSR B1821$-$24) to
16.45 ms (PSR J1022+1001), and their profiles are not only vastly
different in shape and frequency development (KXL98 and XKJ98), but
also dissimilar in their polarization structure (XKJ98).  While, for
instance, in the cases of PSRs J1730$-$2304 and B1821$-$24 a highly
linearly polarised component seems to change in intensity, it is a
weakly polarised component in the case of PSR J1022+1001.
%

An other promiment example where profile changes for a recycled
pulsar have been noticed, are those of the binary pulsar PSR B1913+16
which were described by Weisberg, Romani \&
Taylor~(1989)\nocite{wrt89}, Cordes, Wasserman \&
Blaskiewicz~(1990)\nocite{cwb90} and Kramer (1998)\nocite{kra98b}. The
observed secular, small change in the amplitude ratio and now also
separation of the components is evidently caused by geodetic
precession of the neutron star. However, the time scales of the
profiles changes discussed here are by far shorter and also the
amplitudes involved are dramatically larger.  In combination with the
stable polarization angle swing (at least for PSR J1022+1001), we can
certainly exclude a precession effect for the profile changes of our
studied sample.

The most simple explanation for the observations would be if we had
discovered a mode change as long known for normal pulsars. Although
mode changes are not understood even for slowly rotating pulsars, we
would not have to invoke previously unknown effects. Comparing the
number of normal pulsars known when Backer (1970) \nocite{bac70a}
discovered mode changing, we note that it is similar to the number of
MSPs known now.  However, if the profile changes were just another
aspect of the mode changing for normal pulsars, we would expect
similarly that even in such a case a large number of typically $10^4$
pulses should be sufficient to average out any random fluctuation in
the individual pulses, i.e.~producing a stable waveform.  For PSR
J1022+1001 this would mean to obtain a non-changing pulse profile
already after only 3 min of integration time, in contrast to what is
observed, which is a pulse shape changing {\em smoothly} on much
longer time-scales. Besides, except for the case of PSR J0437$-$4715
reported by Vivekanand et al.~(1998), the pulse shape changes
discussed here seem to appear only in certain parts of the profile
while others are obviously unaffected. This together with the obvious
lack of a relation between pulse shape and intensity is unusual for
the moding behaviour as seen in slowly rotating pulsars.

Most important, however, is that the ``classical'' mode changing does
not provide an explanation for the extraordinary narrow-band variation
of the profile of PSR J1022+1001, which is most reminiscent of a
scintillation pattern. Since we excluded propagation effects caused by
the interstellar medium, the data could be interpreted as a
magnetospheric propagation effect but also in the context of a previously
unnoticed narrow-band property of the emission process. The latter
would be a surprising result since most of the previous pulsar studies
favour a broad band emission process (e.g.~\cite{ls98}). We note here
that Smirnova \& Shabanova (1992) \nocite{ss92} describe simultaneous
observations of PSR B0950+08 at very low frequencies of 60 MHz and 102
MHz. Observing with only one linear polarization they report a
previously unnoticed profile change of this source which does not seem
to occur at both frequencies at the same time. Similar to our
observations, they noticed a narrow-band variation of the pulse
profile at both frequencies with a characteristic bandwidth of 30--40
kHz. Arguing that recording only one linear polarization is not
responsible for this effect, they also consider a narrow band property
of the emission process or a scintillation effect of spatially
separate sources of emission. Smirnova \& Shabanova favour the latter
explanation and give estimates for the separation of the emission
regions. Applying similar calculations to our case, we however easily
derive differences in emission height which are larger than the
light-cylinder radius of PSR J1022+1001.

It is interesting to note that the profile changes of PSR J1022+1001
bear certain similarities to the behaviour of the well known
mode-changing pulsar PSR B0329+54 (\cite{bmsh82}). McKinnon \& Hankins
(1993)\nocite{mh93b} pointed out that ``gated'' pulse profiles of PSR
B0329+54 produced by single pulses sorted according to their
intensity, revealed a shift in the pulse longitude of the core
component depending on its intensity. In order to explain this effect,
they considered a different emission height for strong and weak pulses
as well as a circular motion of the core component around an axis
off-center to the magnetic axis. The profile changes in PSR J1022+1001
could be explained in a similar manner, assuming that a core component
moves, for instance, on an annulus whose center is displaced from the
magnetic axis but closer to the emission region of the leading pulse
peak. Those profiles with an amplitude ratio larger than unity
(cf.~Sect.~\ref{profilechanges}) are then produced when the core
component is positioned in such a way that it adds to the observed
intensity of the first pulse peak. Most of the time, however, it will
be away from the first pulse peak, leading to an average amplitude
ratio lower than unity as observed. Since core components are mostly
associated with circular polarization rather than linear, this simple
picture also provides an explanation why only the circular
polarization is changing whereas the linear remains unchanged. A rough
estimate for the displacement can be calculated by using Eqn.~(5) of
McKinnon \& Hankins (1993), a lower limit for the magnetic inclination
angle $\alpha$ of $60^\circ$ (XKJ98) and the spacing of the centroids
of components 3 and 5 of $\Delta t\sim0.55$ms
(Fig.~\ref{centrhist}). This results in a displacement of $\sim1.2$km,
which corresponds interestingly to the radius of a dipolar polar cap
for PSR J1022+1001. A movement of the core on a circular path would,
however, imply a typical time scale for the profile changes, which is
not observed. If the motion of the core component happens instead in
an irregular manner, obvious time scales might not be present.
Nevertheless, fluctuation spectra of observed single pulses may be
able to resolve a possible movement of the core. Those results should
be frequency independent, since all profile changes should obviously
occur simultaneously over a wide range of frequencies. Single pulse
studies also offer a chance to detect possible correlations between
the intensity of single pulses and the resulting average pulse profile
as in the cases of PSR B0329+54 (\cite{mh93b}) or PSR J0437$-$4715
(\cite{jak+98}). We note that a preliminary analysis of recent Arecibo
data at 430 MHz suggest that ``giant pulses'' for PSR J1022+1001 occur --
if present at all -- much less than once per $10^4$ stellar rotations,
which is already much less than observed for the Crab pulsar
(e.g.~\cite{lcu+95}) or PSR B1937+21 (e.g.~\cite{sb95};
\cite{cstt96}).

Although this above simple picture can apparently explain some of the
observed features at least qualitatively, it bears the fundamental
problem that we still would not know what causes this motion of
individual components. The $E\times B$-drift considered by McKinnon \&
Hankins (1993) would presumably cause a regular motion. Similarly, the
model provides unfortunately no direct explanation of the observed
narrow-band variation of the pulse profile. Actually, if we are
dealing with the same emission mechanism as for normal pulsars (see
KXL98, XKJ98 and Jenet et al.~1998) and if we cannot explain the data
by the known moding behaviour, then we are left with a propagation
effect in the pulsar magnetosphere. This might be combined with
different emission altitudes for different parts of the profile and/or
differential absorption properties of the magnetosphere above the
polar cap.  Indeed, one could interpret the position angle swing of
PSR J1022+1001 as the composition of two separate {\sf S}-swings which
are delayed to each other and thus represent (independent) emission
from different altitudes. In that case, the ``notch'' in the swing
would mark the longitude where the trailing part of the pulse starts
to dominate over the leading one. Applying, however, the model derived
by Blaskiewicz, Cordes \& Wasserman (1991)\nocite{bcw91} to estimate
the emission height based on polarization properties, we would derive
a negative emission altitude for the trailing profile part. More
conventionally we could use the spreads in the centroids, $\Delta t$,
of the fitted Gaussian components as an estimator for a change in
emission height, $\Delta r$. A rough estimate is given by $\Delta r =
c\Delta t/(1+\sin\alpha)$, where $\alpha$ is the magnetic inclination
angle and $c$ the speed of light (see eg.~\cite{mh93b}).  Using the
largest spread as found for component 1 (i.e.~0.082 ms,
cf.~Tab.~\ref{stattab}), and again $\alpha=60^\circ$ (XKJ98), we
derive a change of $\Delta r \sim 130$ km. This value is still smaller
than the light cylinder radius of 785 km.

Although we can apparently construct a simple phenomenological model
which can explain some observations qualitatively, a propagation
effect in the pulsar magnetosphere might be still the most probable
explanation for the observed phenomena. In conclusion, we believe that
this interpretation and the reason for the observed narrow-band
variation of the pulse shape should be addressed with future
simultaneous multi-frequency observations of these interesting
sources. Only such observations have the potential to distinguish
between a propagation effect in the pulsar magnetosphere, which can be
expected to be frequency dependent, and those involving a reformation
of the emitting regions, which should produce frequency independent
properties.

\section{Summary}
\label{summary}

Focussing in particular on PSR J1022+1001, we have demonstrated that a
sample of MSPs shows distinct and unusual profile changes.  We argued
that these profile changes are not caused by instrumental effects or
represent a propagation effect in the interstellar or interplanetary
medium. In fact, we conclude that the observed variations in pulse
shapes (in time and frequency) are intrinsic to the pulsars and that
they are {\em not} consistent with the mode changing effect known for
normal pulsars.

We have shown that the profile changes can have a significant impact
regarding the apparent timing stability of MSPs. We suggest the usual
template matching procedure to be extended by allowing for variations
of the amplitudes of different profile component. As demonstrated for
PSR J1022+1001 this procedure improves the timing accuracy
significantly and has led to the first proper motion measurement for
this pulsar.

\acknowledgments 

We are indebted to all people involved in the project to monitor
millisecond pulsars in Effelsberg, in particular to Axel Jessner and
Alex Wolszczan. MK acknowledges the receipt of the Otto-Hahn Prize,
during whose tenure this paper was written, and the warm hospitality
of the Astronomy Department at UC Berkeley. FC is a Marie Curie Fellow.

\newpage


\newpage

\newpage

\begin{table}
\caption{\label{stattab}
Statistical results of fitting a model of five Gaussian components to
profiles observed at 1410 MHz. In the first case (second and third
column) only the relative amplitudes were adjusted, while in the second
also the Gaussian centroids were varied (fourth and fifth column).
For the amplitudes, which are quoted in units of the trailing pulse
peak, we list variances and their ratio to the mean amplitudes.
For the centroids of the components in units of milliseconds (relative
to the fiducial point), we quote variances and their values normalized
to the (fixed) width of each component.}

\begin{center}
\begin{tabular}{lccccc}
\hline\hline
 & \multicolumn{2}{c}{Amplitudes} & & \multicolumn{2}{c}{Centroids} \\
 \cline{2-3} \cline{5-6} 
Comp & $\sigma$ & $\sigma$/mean & & $\sigma$ (ms) & $\sigma$/width \\
\hline
I.   & 0.027 & 0.128 & & 0.082 & 0.053 \\
II.  & 0.059 & 0.134 & & 0.057 & 0.059 \\
III. & 0.090 & 0.160 & & 0.017 & 0.031 \\
IV.  & 0.052 & 0.122 & & 0.014 & 0.039 \\
V.   & 0.042 & 0.043 & & 0.004 & 0.009 \\
\hline
\end{tabular}
\end{center}

\end{table}

\newpage

\begin{table}
\caption{\label{timpar} Timing parameters of PSR~J1022+1001.
Numbers in parentheses are $1\sigma$ uncertainties
derived from a combination of all four available data sets.}

\begin{center}
\begin{tabular}{ll}
\hline\hline
Ecliptic Longitude,  $\lambda$ (deg)  \dotfill & 153.8659226(8)   \\
Ecliptic Latitude,  $\beta$ (deg)   \dotfill & $-$0.0641(1)   \\
Proper Motion in $\lambda$, $\mu_\lambda$ (mas/yr) \dotfill 
                                              & $-$17(2) \\
Period (ms)                          \dotfill & 16.4529296832030(4) \\
Period derivative ($10^{-20}$)       \dotfill & 4.341(4) \\
Epoch of period (MJD)                \dotfill & 50250 \\
Dispersion measure\tablenotemark{a} (pc\,cm$^{-3}$)  \dotfill 
                                              & 10.246 \\
Projected semi-major axis (light s)  \ldots   & 16.765409(2)    \\
Eccentricity                         \dotfill & 0.00009735(8)   \\
Time of periastron passage (MJD)     \dotfill & 50246.716(2)    \\
Orbital Period (days)                \dotfill & 7.805130162(6)  \\
Angle of periastron (degrees)        \dotfill & 97.67(7)        \\
\noalign{\smallskip}
Right Ascension\tablenotemark{b}     \dotfill & 
                                $10^{\rm h}22^{\rm m}57\fs997(9)$ \\
Declination\tablenotemark{b}         \dotfill &  
                                $10^\circ 01'52\farcs1(3)$ \\
\hline
\end{tabular}
\tablenotetext{a}{Held fixed}
\tablenotetext{b}{Calculated from $\lambda$ and $\beta$}

\end{center}

\end{table}

\end{document}